\documentclass[useAMS,usenatbib]{mn2e}
\usepackage{epsfig}
\usepackage{graphicx}
\usepackage{epstopdf}

\def\simgt{\lower.5ex\hbox{$\; \buildrel > \over \sim \;$}}
\def\simlt{\lower.5ex\hbox{$\; \buildrel < \over \sim \;$}}
\def\rtwopf{{\it r2p5}}
\def\rfive{{\it r5}}
\def\rten{{\it r10}}
\def\rtwentyf{{\it r25}}
\def\rtwentyfisol{{\it r25isol}}

\def\rhratio{R_{h,FG}/R_{h,SG}}

\def\nratio{N_{SG}/N_{FG}}
\def\nratioglob{(N_{SG}/N_{FG})_{glob}}

\newcommand\aj{{AJ}}%
\newcommand\apj{{ApJ}}%
\newcommand\aap{{A\&A}}%
\newcommand\mnras{{MNRAS}}%
\title[Multiple population dynamics]{Dynamical Evolution and Spatial
  Mixing of Multiple Population Globular Clusters}
\author[E. Vesperini et al.]  {Enrico Vesperini$^1$, Stephen L.W. McMillan$^2$, Francesca D'Antona$^3$, Annibale D'Ercole$^4$\\
$^1$Department of Astronomy, Indiana University, Bloomington, IN 47405, USA\\
$^2$Department of Physics, Drexel University, Philadelphia, PA 19104, USA\\
$^{3}$INAF- Osservatorio Astronomico di Roma, via di Frascati 33,
  I-00040 Monteporzio (Italy)\\
$^{4}$INAF- Osservatorio Astronomico di Bologna, via Ranzani 1,
  I-40127 BOLOGNA}
\begin{document}
\maketitle

\label{firstpage}

\begin{abstract}
  Numerous spectroscopic and photometric observational studies have
  provided strong evidence for the widespread presence of multiple
  stellar populations in globular clusters.  In this paper we study
  the long-term dynamical evolution of multiple-population clusters,
  focusing on the evolution of the spatial distributions of the first-
  (FG) and second-generation (SG) stars.  In previous studies we have
  suggested that SG stars formed from the ejecta of FG AGB stars are
  expected initially to be concentrated in the cluster inner regions.
  Here, by means of $N$-body simulations, we explore
  the time scales and the dynamics of the spatial mixing of the FG and
  the SG populations and their dependence on the SG initial
  concentration.

  Our simulations show that, as the evolution proceeds, the radial
  profile of the SG/FG number ratio, $\nratio$, is characterized by
  three regions: 1) a flat inner part; 2) a declining part in which FG
  stars are increasingly dominant; and 3) an outer region where the
  $\nratio$ profile flattens again (the $\nratio$ profile may rise slightly
  again in the outermost cluster regions).  Until mixing is complete
  and the $\nratio$ profile is flat over the entire cluster, the
  radial variation of $\nratio$ implies that the fraction of SG stars
  determined by observations covering a limited range of radial
  distances is not, in general, equal to the SG global fraction,
  $\nratioglob$.  The distance at which $\nratio$ equals $\nratioglob$
  is approximately between 1 and 2 cluster half-mass radii.  The time scale for
  complete mixing depends on the SG initial concentration, but in all
  cases complete mixing is expected only for clusters in advanced
  evolutionary phases, having lost at least 60--70 percent of
  their mass due to two-body relaxation (in addition to the early
  FG loss due to the cluster expansion triggered by SNII ejecta and
  gas expulsion).The results of our simulations suggest that in many Galactic globular clusters the SG should still be more spatially concentrated than the FG. 
\end{abstract}

\begin{keywords}
globular clusters:general, stars:chemically peculiar, methods:N-body
simulations 
\end{keywords}

\section{Introduction}
\label{sec:intro}
An increasing number of spectroscopic and photometric observational
studies have provided strong evidence for a widespread presence of
multiple stellar populations in globular clusters.  The observed
star-to-star variations of light elements, such as Na, O, Al, and Mg,
indicates that a significant fraction (50-80\%) of globular cluster
stars must have formed out of matter processed through a
high-temperature CNO cycle in a first generation of stars (hereafter
FG; see e.g. Carretta et al. 2009a, 2009b and references therein).
Photometric studies, by revealing the presence of multiple main
sequences, subgiant, and red-giant branches in numerous clusters, have
buttressed the spectroscopic evidence and added important new elements
to the observational framework of multiple population in globular
clusters. In particular, photometric measurements have shown that some
clusters host a population of very He-rich stars among the
second-generation (hereafter SG) population of some clusters (see
e.g. Piotto et al. 2007; see also D'Antona et al. 2002, D'Antona \&
Caloi 2004, D'Antona \& Caloi 2008).
Recently, the first direct measurement of He abundances of two stars
in NGC 2808 has confirmed the strong He-enhancement suggested by those
photometric studies (Pasquini et al. 2011; see also Dupree et al. 2011
for evidence of He enhancement in $\omega$ Cen stars).

Models for the source of polluted gas from which
SG stars formed include AGB stars (see e.g. Cottrell \& Da Costa 1981,
Ventura et al. 2001), rapidly rotating massive stars (Decressin et
al. 2007), and massive binary stars (De Mink et al. 2009).  
A few studies have also explored some aspects of cluster formation and
evolution for some of these models (see e.g. D'Ercole et al. 2008,
2010, 2012, Bekki 2011, Decressin et al. 2008, 2010; see Renzini 2008
and Gratton et al. 2012 and references therein for a review).  

In D'Ercole et al. (2008), we focused our attention on the AGB model
and explored the formation and dynamical evolution of multiple
populations by means of hydrodynamical and N-body simulations.  Our
simulations show that the AGB ejecta form a cooling flow and rapidly
collect in the innermost regions of the cluster, forming a
concentrated SG stellar subsystem (see also Bekki 2011).  In order to
form the number of  
SG stars observed today, the FG cluster must have been considerably
more massive than it is now. The N-body simulations presented in
D'Ercole et al. (2008) show that the early expansion triggered by the
loss of mass in the form of SNII ejecta leads to a strong preferential
loss of FG stars resulting in a cluster in which the number of SG
stars is similar to (or even larger than) that of FG stars as observed
in several Galactic globular clusters (see e.g. Carretta et al. 2009a,
2009b).

In subsequent papers we expanded the initial models presented in
D'Ercole et al. (2008) to further explore the origin of the observed
abundance patterns (D'Ercole et al. 2010, 2012), the connection between
multiple population globular clusters and the Galactic stellar halo
(Vesperini et al. 2010) and the implications for the disruption of FG
and SG binary stars (Vesperini et al. 2011).

In this paper we focus our attention on the long-term dynamical
evolution of the multiple-population cluster, starting immediately
after the early evolutionary stages during which a large fraction of
the FG population is lost.  We emphasize that, throughout this paper,
when we discuss cluster mass loss we are referring only to mass lost
by the cluster during its long-term relaxation-driven evolution, not
the early phase responsible for the loss of most of the initial FG
population. 

After the early loss of FG stars (e.g. at $\approx 1-2$ Gyr for the simulations presented in D'Ercole et al. 2008), a multiple-population cluster will
start its long-term evolution driven by two-body relaxation with a
similar number of SG and FG stars but with the SG population still
concentrated in the cluster inner regions.  A number of observational
studies (Bellini et al. 2009, Carretta et al. 2010a, Lardo et al. 2011,
Kravtsov et al., 2010, 2011, Nataf et al. 2011, Johnson \& Pilachowski
2012, Milone et al. 2012)  
have found that in several clusters SG stars are indeed preferentially
located in the inner regions and retain some memory of the initial
segregation of the SG population predicted by the models we presented
in D'Ercole et al. (2008).

Understanding the dynamics of the spatial mixing, and specifically the
extent to which memory of the initial SG segregation predicted by the
formation models is retained after one Hubble time of
relaxation-driven evolution, is an essential step to properly
interpreting observational data, as well as testing the key elements
of theoretical scenarios of cluster formation and evolution.

In this paper, by means of N-body simulations, we explore the
structural evolution of multiple population clusters.  We focus our
attention on the spatial mixing of the SG and the FG populations, and
the evolution of the relative spatial distribution of the SG and the
FG stars.  The structure of the paper is as follows: in Section 2 we describe the
initial conditions of our N-body simulations; in Section 3 we present
our results; and in Section 4 we discuss our results and summarize our
main conclusions.

\section{Method and Initial conditions}
\label{sec:initial}
The study presented in this paper is based on N-body simulations run
with the {\tt starlab} package (Portegies Zwart et al. 2001) and
accelerated by GRAPE-6 special purpose hardware (Makino
et al. 2003) and GPU (Gaburov et al. 2009).  We explore the evolution of four systems with different
degrees of initial concentration of the SG population.  
In all cases the initial conditions are set by simply combining
two single-mass King (1966) models: the FG cluster is modeled as a King
model with dimensionless central potential $W_0=7$; the SG system is also
modeled as a King model with $W_0=7$, but it is initially entirely
contained within the inner regions of the FG system.
Our models have an initial ratio of the FG to the SG half-mass radius,
$R_{h,FG}/R_{h,SG}$ equal to $2.5, 5, 10$, and $25$ ($R_{h,FG}$ and $R_{h,SG}$ are the 3D half-mass radii).  Hereafter we
refer to these simulations as \rtwopf, \rfive, \rten, and \rtwentyf.
The models
presented do not include primordial binaries or stellar evolution.
The combined system is initially scaled to a virial ratio $T/V = 0.5$ and is
allowed to dynamically settle into an equilibrium state during the first few
dynamical times of each simulation.

All simulations start with a total number of particles $N=10,000$;
initially the numbers of FG and SG particles are equal.  We assume
that the cluster is tidally truncated and therefore that the King
truncation radius is equal to the cluster Jacobi radius, $R_J$.
Particles are removed from the simulation after reaching a radius
equal to twice the Jacobi radius.  As discussed in Section
\ref{sec:intro}, the initial conditions adopted for our simulations
refer to the possible structural properties of a multiple-population
cluster after the early loss of FG stars.  

To explore the role of the relaxation-driven evaporation of stars on
the cluster structural and mixing properties, we also include a
simulation with the same initial conditions as those of the
{\rtwentyf} system, but without a tidal field or tidal truncation, and
in which stars were removed only after reaching a distance of about
1000 times the cluster half-mass radius.  Hereafter we refer to this
simulation as \rtwentyfisol.

All simulations were run until the total mass of the system was 20
percent or less of its initial value.  For the {\rtwentyfisol} system,
mass loss proceeds at a much slower rate and the simulation was
stopped after the system had lost 20 percent of its initial mass.  In
order to explore the dependence of our results on the number of
particles, two additional simulations with $N=30,000$ and $N=60,000$
particles were also run, with initial structural properties identical
to those of the {\rten} system.

\begin{figure*}
\vbox to220mm{\vfil
\centering{
\includegraphics[width=6cm]{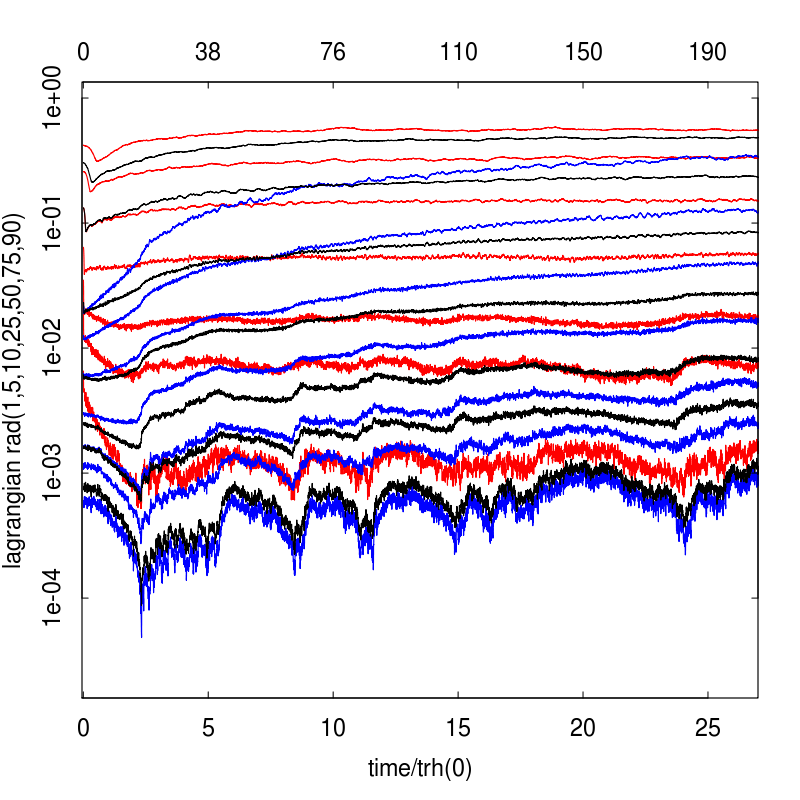}
\includegraphics[width=6cm]{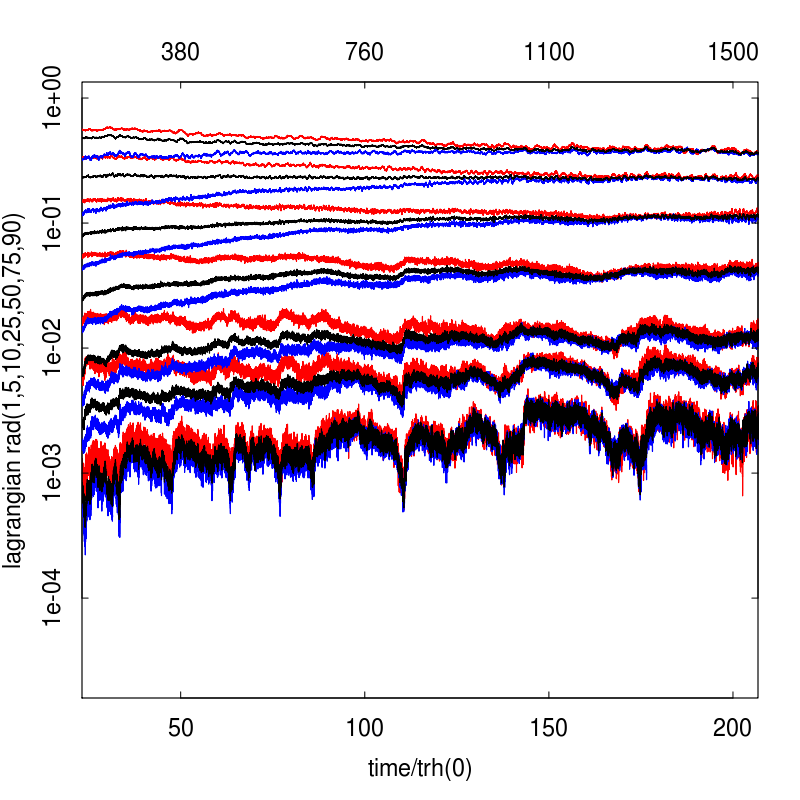}
}
\centering
{
\includegraphics[width=6cm]{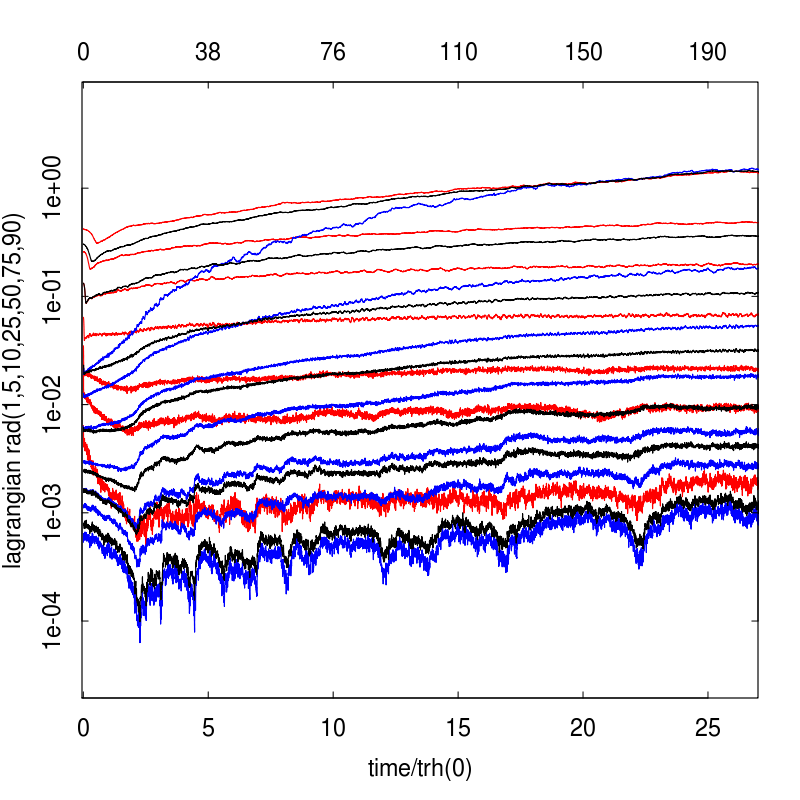}
\includegraphics[width=6cm]{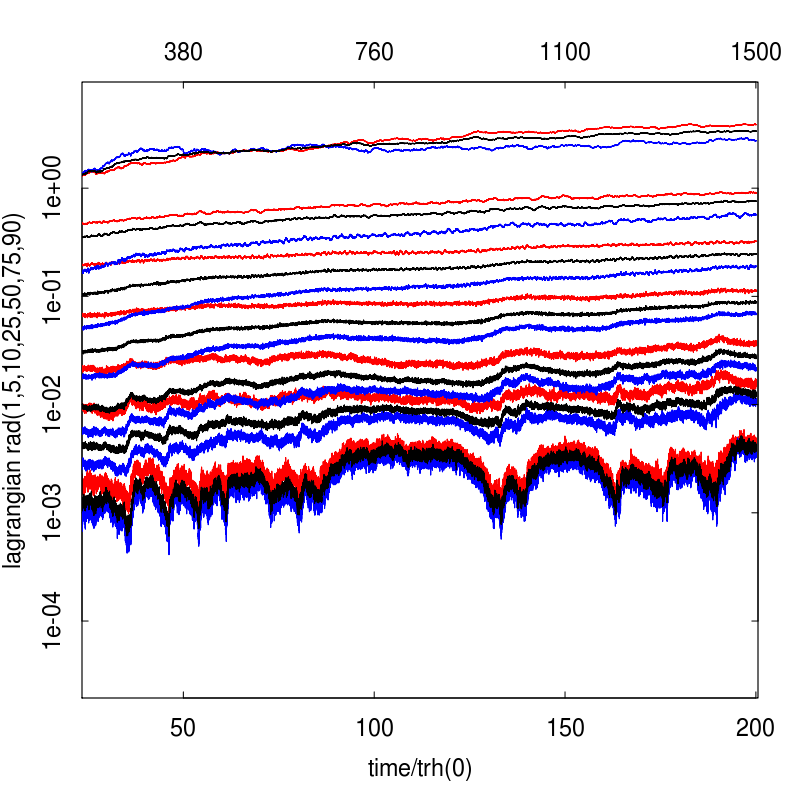}
}
\centering{
\includegraphics[width=6cm]{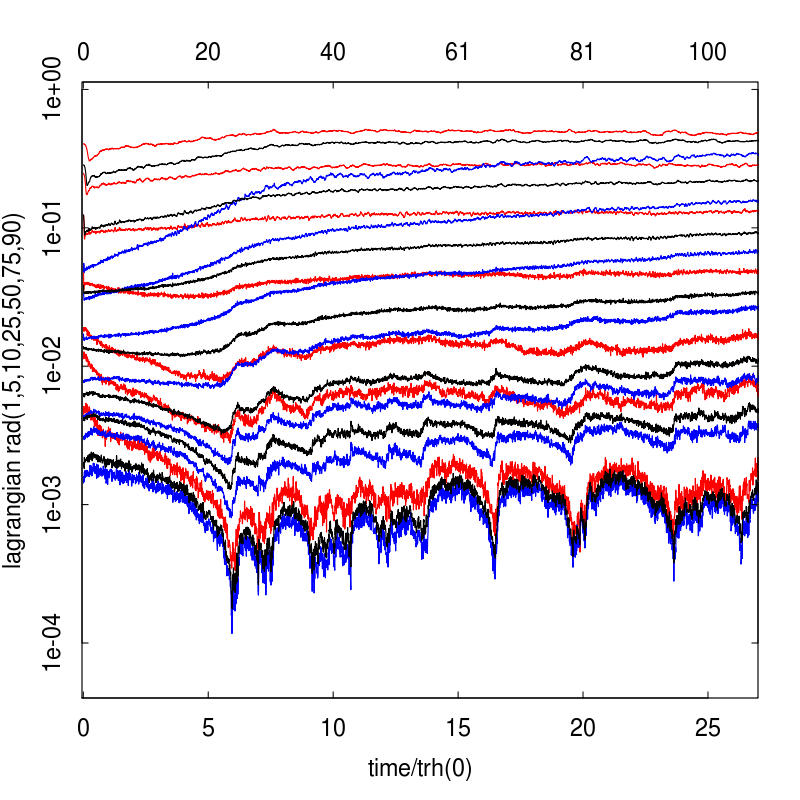}
\includegraphics[width=6cm]{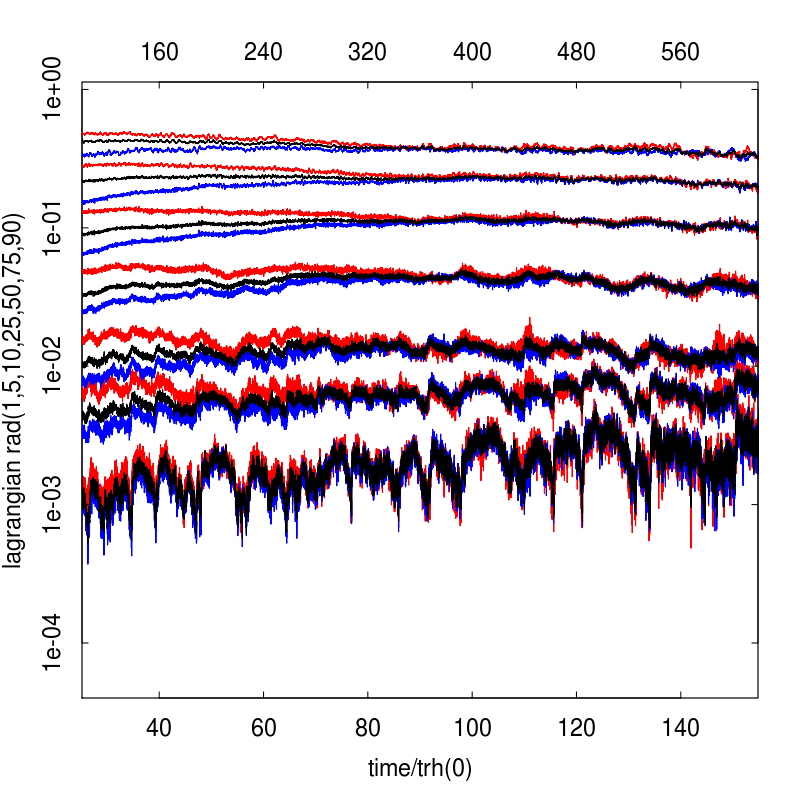}
}
\caption{Time evolution of the 1\%, 5\%, 10\%, 25\%, 50\%, 75\%, 90\%
  lagrangian radii for SG stars (blue lines), FG stars (red lines) and
  all stars (black lines).  The five rows of figures refer (from top
  to bottom) to the \rtwentyf, \rtwentyfisol, \rten, \rfive, and
  {\rtwopf} simulations. In each panel, time is expressed relative
  to the initial half-mass relaxation time of the entire system in the
  bottom horizontal axis, and to the initial half-mass relaxation time
  of the SG subsystem in the upper horizontal axis.  In order to show
  more clearly the details of the lagrangian radius evolution, for
  each simulation (except for \rtwopf) the left panels show only the
  evolution until $\sim 25 t_{rh}(0)$ and the right panels the rest of
  the evolution until the two populations are mixed.}
\label{fig:lagrfig}
\vfil}
\end{figure*}
\begin{figure*}
\centering{
\includegraphics[width=6cm]{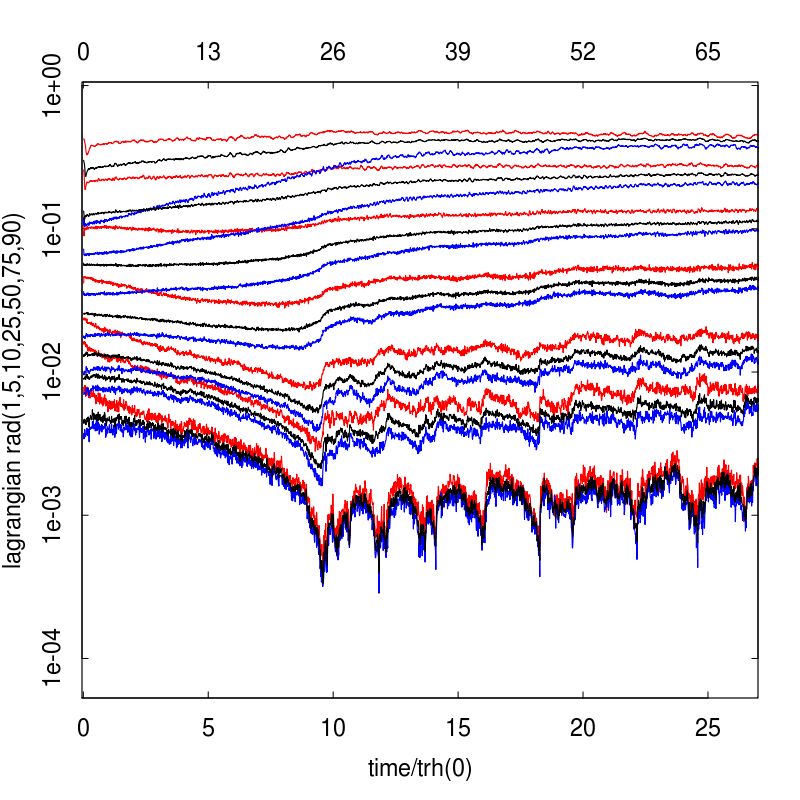}
\includegraphics[width=6cm]{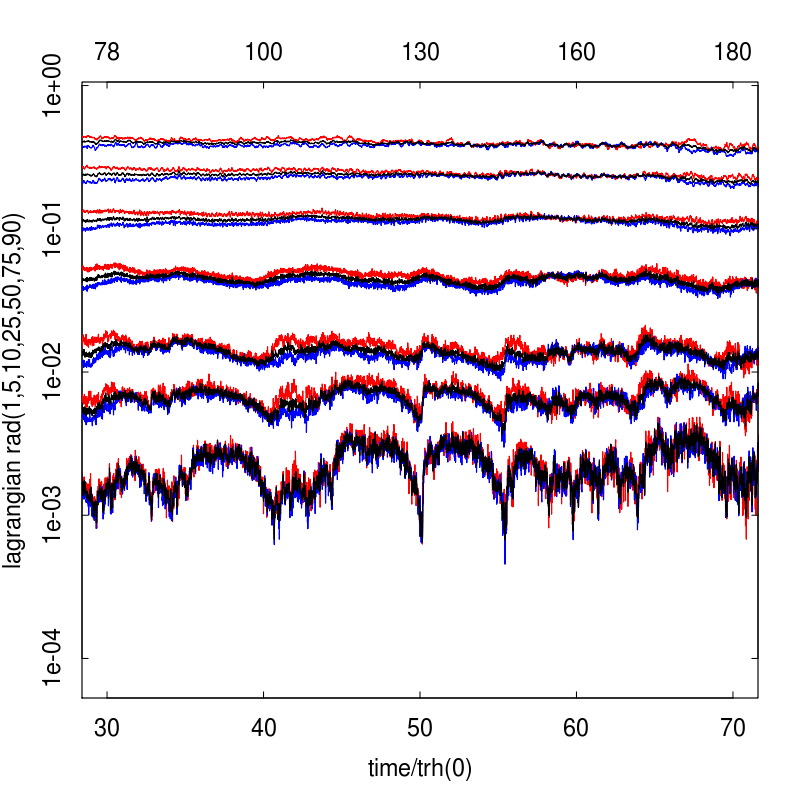}
}
\centering{
\includegraphics[width=6cm]{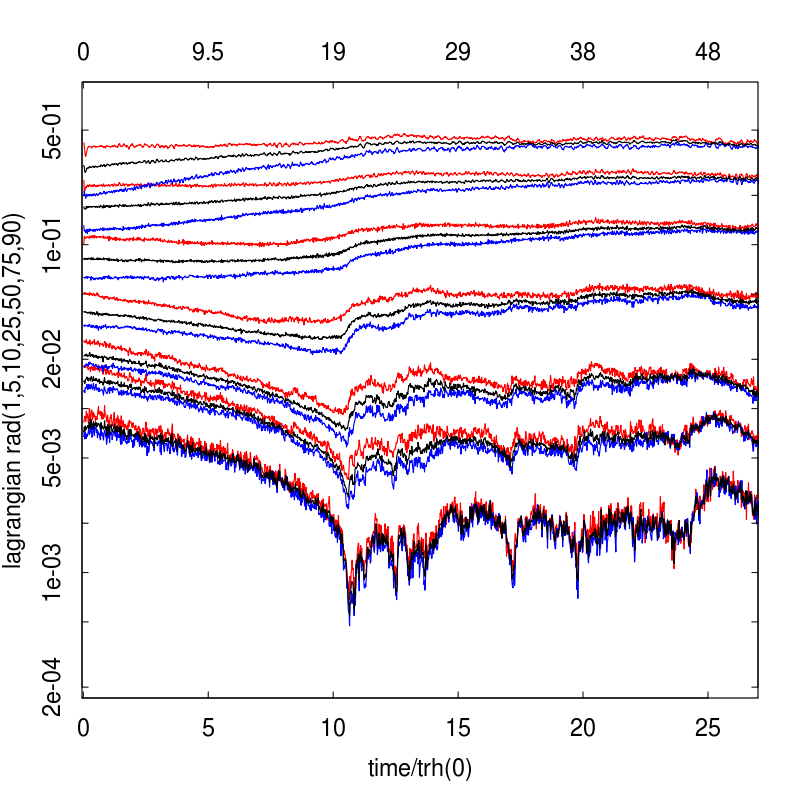}
}

\contcaption{}

\end{figure*}
    
\section{Results}
\subsection{Dynamics of spatial mixing}
We start the presentation of our results by focussing our attention
on the dynamics of the spatial mixing process. We emphasize here that
the simulations presented in this paper are aimed at understanding
the fundamental aspects of the dynamics of multiple-population
clusters; additional simulations currently in progress, including a
spectrum of masses will allow us to more closely compare our
theoretical results with observations.  
The panels of Fig.\ref{fig:lagrfig} show the time evolution of the
Lagrangian radii of the FG and SG populations (along with those of the
entire system), and illustrate the system's structural evolution and
degree of spatial mixing.  Overlap of all Lagrangian radii for the two
populations indicates complete mixing.  As indicated in
Fig.\ref{fig:lagrfig}, the number of initial half-mass relaxation
times (of the whole system or of the SG subsystem) needed to reach
complete mixing of the two populations depends strongly on the initial
concentration of the SG subsystem.

\begin{figure}
\centering{
\includegraphics[width=8cm]{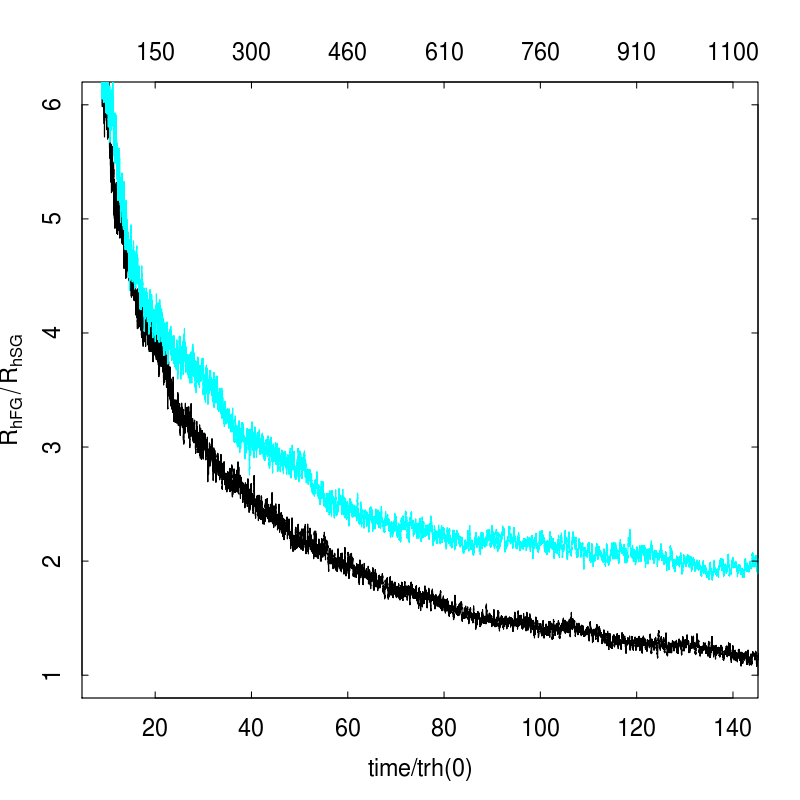}
}
\caption{Time evolution of the ratio of the FG to the SG half-mass
  radii for the {\rtwentyf} (black lower line) and the {\rtwentyfisol}
  (cyan upper line) systems.  Time is normalized to the initial half-mass
  relaxation time of the whole system in the lower axis, and to the
  initial half-mass relaxation time of the SG subsystem in the upper
  axis.}
\label{fig:rhr25}
\end{figure}

To illustrate the dynamics behind the mixing process and the role of
relaxation-driven mass loss, Fig.~\ref{fig:rhr25} compares the time
evolution of the ratio of the FG to the SG half-mass radii,
$\rhratio$, a convenient measure of global mixing, for the {\rtwentyf}
and the {\rtwentyfisol} runs.  Initially, the ratio decreases
relatively rapidly, as two-body relaxation acts to erase the initial
spatial differences between the two populations.
However, its rate of change slows as the system evolves.  The reason
for this is that, after reaching core collapse (at $t \sim 17
t_{rh,SG}(0)$ for both the {\rtwentyf} and the {\rtwentyfisol}
systems), the system enters its post-core collapse expansion phase.
If the cluster did not lose a significant amount of mass during the
expansion, its relaxation time would increase, with the half-mass
radius and half-mass relaxation time growing with time as $R_h \sim
t^{2/3}$ and $t_{rh} \sim t$ (see e.g. Spitzer 1987).  We note that
the same behavior, with the same implications for the mixing process,
is expected no matter what mechanism (e.g. three-body binaries,
primordial binaries, stellar evolution mass loss) drives the cluster
expansion.  (See e.g. Gieles et al. 2011 for a recent discussion of
the role of mass loss due to stellar evolution in providing the energy
needed to balance the energy flow from the cluster half-mass radius.)

The increasing relaxation time means that the evolution of any cluster
property driven by two-body relaxation slows down accordingly.  If we
define the cluster dynamical age as $\tau=\int_0^t\hbox{d}t/t_{rh}(t)$
and assume $t_{rh}\sim t$, it follows that, during this phase, $\tau
\sim \log\,t$ and the rate of the cluster dynamical aging slows down
as $\hbox{d}\tau/\hbox{d}t \sim 1/t$.  The decreasing rate of spatial
mixing of the FG and SG populations evident in Fig.~\ref{fig:rhr25} is
a manifestation of the decreasing cluster aging rate.

In reality, clusters lose mass, and the time evolution of $t_{rh}$ and
$\tau$ deviate from the expressions derived on the assumption of zero
mass loss.  While the spatial mixing rate decreases with time for both
the {\rtwentyf} and the {\rtwentyfisol} systems, Fig.~\ref{fig:rhr25}
shows that the evolution of $\rhratio$ for the {\rtwentyfisol} system
is significantly slower than for the {\rtwentyf} system.  This
difference is due to the different mass loss rates of the two systems.
As the amount of mass lost increases, the growth of the cluster
relaxation time gradually slows, and eventually it starts to decrease,
both as a result of the decreasing remaining mass and of the eventual
contraction of the half-mass radius.  As the cluster mass decreases so
does its Jacobi radius, and the outer, less relaxed (and less mixed),
layers are gradually stripped away.  Thus, by enhancing the loss of
less-mixed layers and slowing the growth of the cluster relaxation
time, mass loss accelerates the cluster dynamical aging process and
the evolution toward complete spatial mixing.

\begin{figure}    
\centering{
\includegraphics[width=8cm]{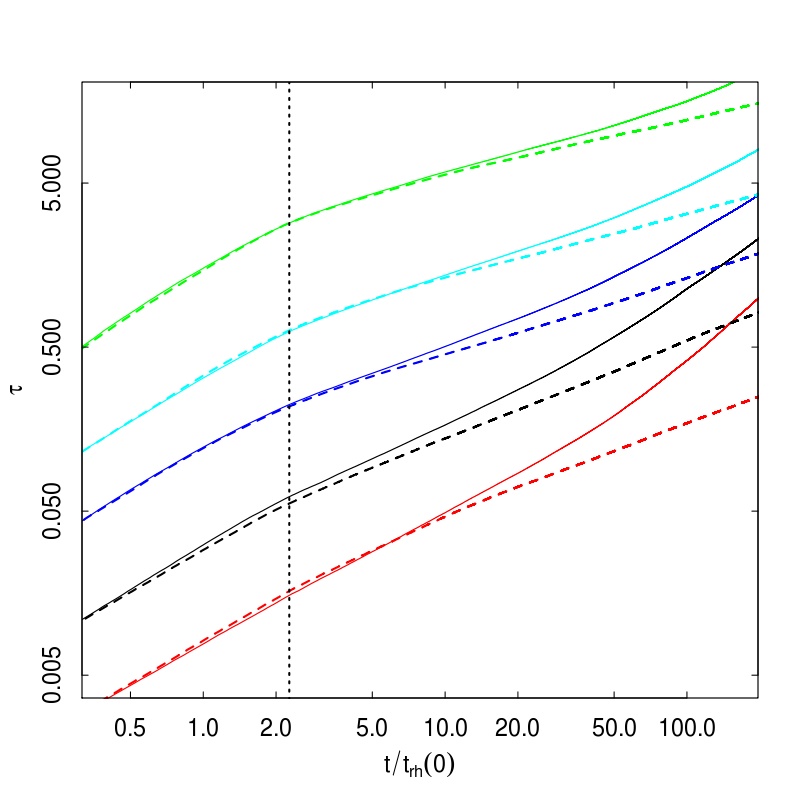}
}
\caption{Time evolution of the cluster dynamical age $\tau$ (see text
  for definition) for the {\rtwentyf} (solid lines) and the
  {\rtwentyfisol} (dashed lines) systems. The five lines for each
  system refer (from top to bottom) to the values of $\tau$ measured
  at the 25\%, 40\%, 50\%, 60\%, 75\% lagrangian radii. The vertical
  dotted line marks the time of core collapse (see also the top left
  panel of Fig.~\ref{fig:lagrfig}). }
\label{fig:tau}
\end{figure}

We further illustrate these points in Fig.~\ref{fig:tau}.  Here we
focus on the local relaxation time $t_{relax}(r) = 0.34
\sigma(r)^3/(G^2 m \rho(r) \ln (0.11 N))$, which provides a more
accurate measure of the level of mixing expected at different
distances from the cluster center.  Here, $\sigma(r)$ is the cluster
1-D velocity dispersion, $\rho(r)$ the cluster mass density and $N$ is
the total number of stars.  Fig.~\ref{fig:tau} shows the time
evolution of the local cluster dynamical age $\tau(r,t) =
\int_0^t\hbox{d}t/t_{relax}(r,t)$ at different distances from the
cluster center.  The slowdown in the time evolution of $\tau$ when the
cluster expansion starts (marked by the vertical dotted line) and the
differences between the dynamical aging of the {\rtwentyf} and the
{\rtwentyfisol} systems are evident in this figure.

Fig.~\ref{fig:tau} also shows the radial dependence of the cluster
dynamical age: as expected, $\tau(r,t)$ decreases at larger distances
from the cluster center, and the cluster outer regions mix on longer
time scales than the inner regions.  The implications of this radial
variation of the dynamical age on the mixing process, and its imprint
on the radial variation of the fraction of SG stars, are discussed in
more detail in Section \ref{sec:radial}.

The general evolution toward spatial mixing is therefore driven by
internal two-body relaxation and accelerated by mass loss.  The mass
loss rate due to two-body relaxation is determined primarily by the
strength of the external tidal field (see e.g. Vesperini \& Heggie
1997, Baumgardt \& Makino 2003, Gieles \& Baumgardt 2008), while the local internal relaxation
rate is determined by the cluster structural properties and their
radial variation within the cluster.  The different processes (and
time scales) involved in the mixing dynamics imply that, in general, it
is not possible to determine a universal mixing time scale simply in
terms of the cluster initial half-mass relaxation time scale.  
The {\rtwentyf} and {\rtwentyfisol} systems share the
same initial structure but, after a given number of initial half-mass
relaxation time scales, because of their different mass loss history,
they have reached a significantly different degree of mixing.

\subsection{Dependence on the SG initial concentration}
\label{sec:concent}
We have carried out a set of four simulations with different SG
initial spatial concentrations.  In comparing the evolution of these
systems, we assume they represent clusters with the same initial
masses and tidal radii (equal to the Jacobi radius), but different
internal structures determined by the concentration of the SG
population.
\begin{figure}    
\centering{
\includegraphics[width=7.5cm]{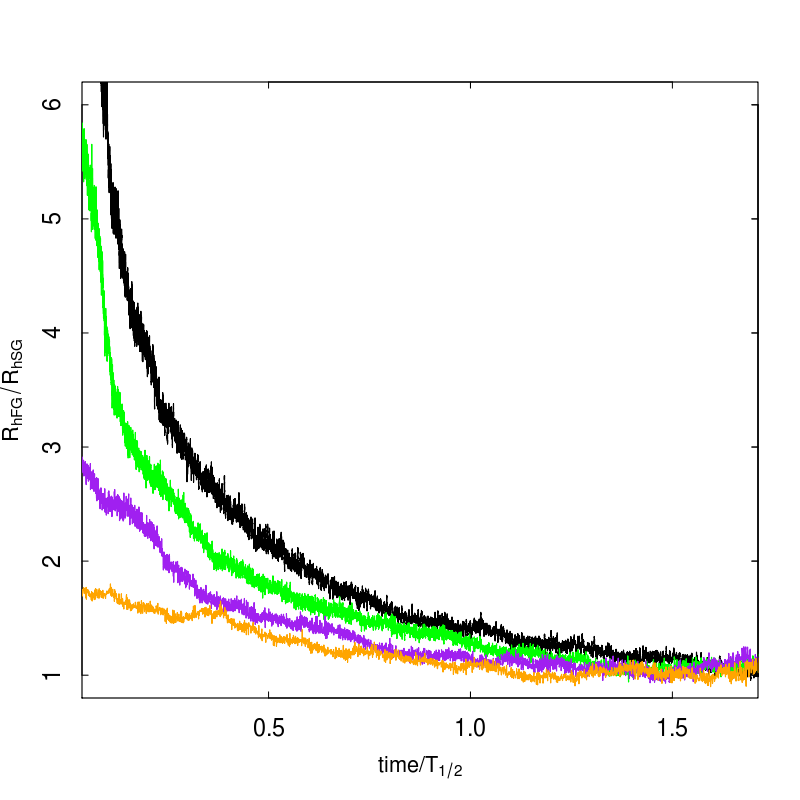}
}
\centering{
\includegraphics[width=7.5cm]{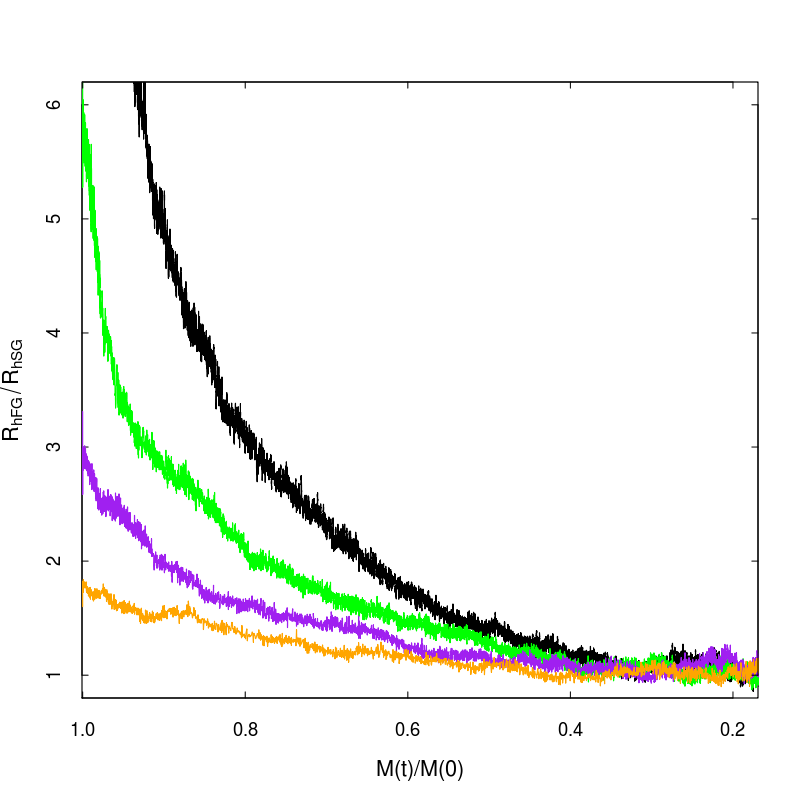}
}
\centering{
\includegraphics[width=7.5cm]{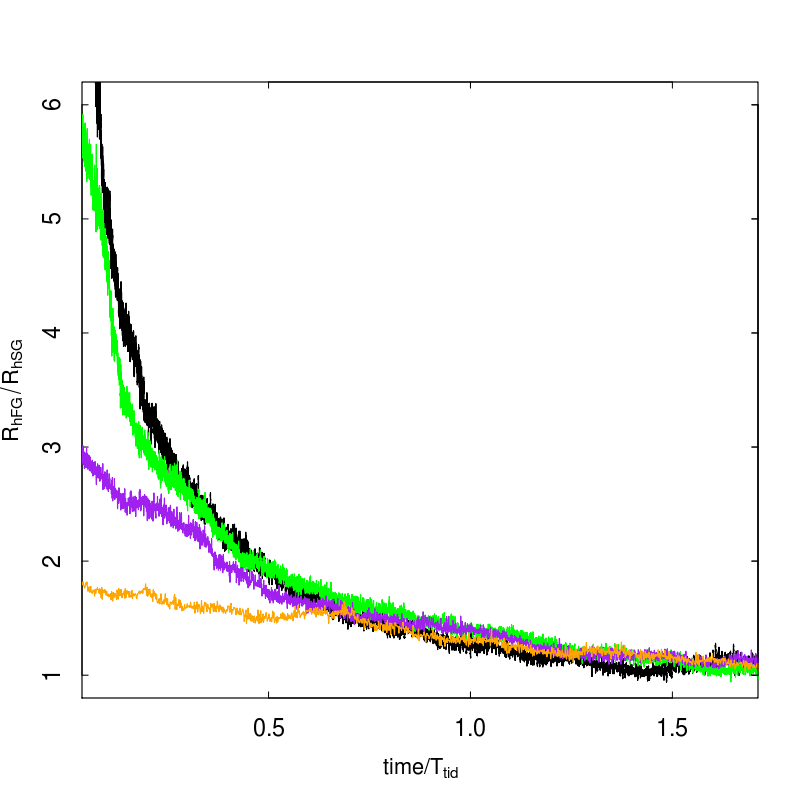}
}
\caption{Evolution of $\rhratio$ as a function of time normalized to
  the half-mass lifetime of each cluster, $T_{1/2}$ (top panel), the
  ratio of the cluster mass to the initial mass (middle panel), and
  time normalized to a common reference time $T_{tid} =
  0.1N^{0.65}/\omega$ (lower panel; see Section \ref{sec:concent} for
  a discussion of $T_{tid}$).  The lines in each panel refer to the
  {\rtwentyf} (black), {\rten} (green), {\rfive} (purple), and the
  {\rtwopf} (orange) systems.}
\label{fig:rhrall}
\end{figure}
As shown in numerous investigations based on Fokker-Planck and N-body
simulations (see e.g. Chernoff \& Weinberg 1990, Vesperini \& Heggie
1997, Baumgardt \& Makino 2003, Gieles \& Baumgardt 2008), systems
with the same tidal radius and mass share a similar relaxation-driven
dissolution time scale, with only a weak dependence on the internal
structure of the cluster (see e.g. Gieles \& Baumgardt 2008).  The
four systems investigated here span almost an order of magnitude in
initial half-mass relaxation time, but only a factor of two in the
half-mass lifetime (defined as the time needed to lose half of their
initial mass due to the effects of two-body relaxation).
 The upper and middle panels of
Fig.~\ref{fig:rhrall} show the evolution of $\rhratio$ as a function
of time normalized to each cluster's half-mass time scale, $T_{1/2}$,
and as a function of the cluster total mass (normalized to its initial
value).

The lower panel of Fig.~\ref{fig:rhrall} shows $\rhratio$ as a
function of time normalized to a common reference time, $T_{tid}$,
defined as $T_{tid}=0.1N^{0.65}/\omega$, where
$\omega=\sqrt{GM/3R_J^3}$.  The
definition of $T_{tid}$ follows from studies showing that this
functional form provides a good fit to the scaling of the
relaxation-driven tidal dissolution (or half-mass) time scale with
initial cluster properties.  The exact value of the numerical factor
needed for $T_{tid}$ to match half-mass time scale depends on the
cluster mass spectrum and the cluster structure (see e.g. Gieles \&
Baumgardt 2008 for a discussion on the dependence on the latter).
Note that $T_{tid}$ is closely related to the dynamical family
parameter introduced by Chernoff \& Weinberg 1990 (see their Table 3).

\begin{figure}    
\centering{
\includegraphics[width=8.5cm]{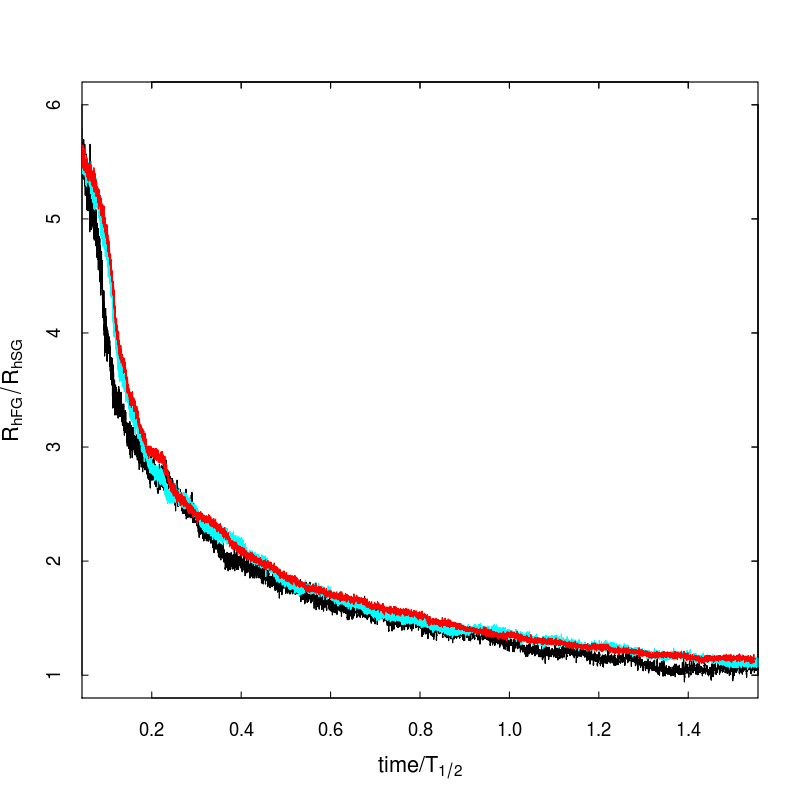}
}
\caption{Evolution of $\rhratio$ as a function of time normalized to
  $T_{1/2}$ for the {\rten} system with (from top to bottom)
  $N=60,000$ (top red line), $N=30,000$ (middle cyan line), and
  $N=10,000$ (lower black line).}
\label{fig:ndepend}
\end{figure}

\begin{figure}    
\centering{
\includegraphics[width=8.5cm]{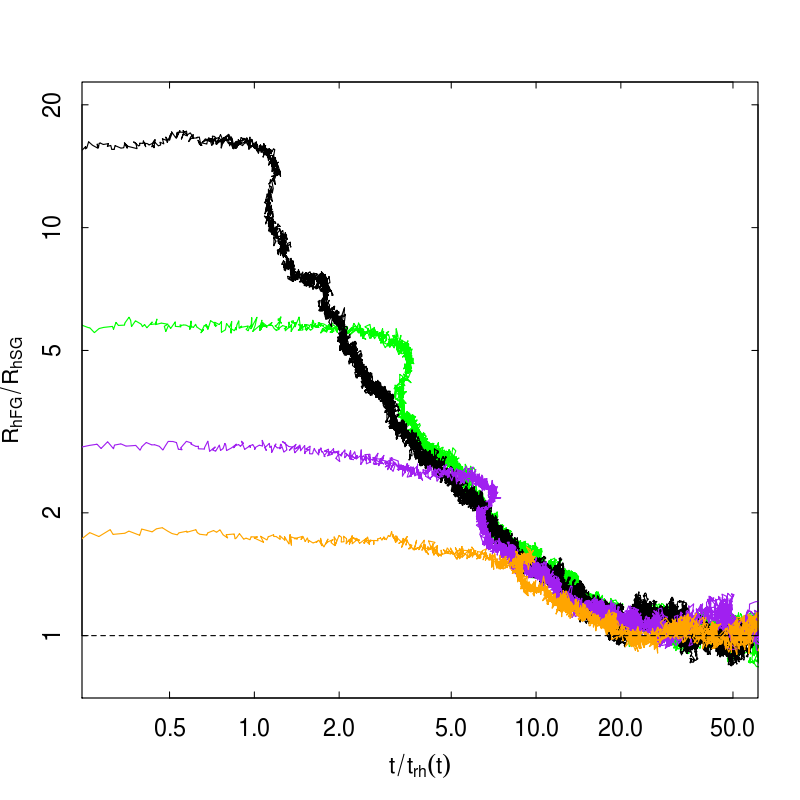}
}
\caption{Evolution of $\rhratio$ as a function of time $t$ relative to
  the instantaneous cluster half-mass relaxation time scale
  $t_{rh}(t)$, calculated using the $3D$ half-mass radius, $R_h$. The lines refer to the
  {\rtwentyf} (black), {\rten} (green), {\rfive} (purple), and the
  {\rtwopf} (orange) systems.} 
\label{fig:rhrall2}
\end{figure}

As just discussed, both internal two-body relaxation and (indirectly)
mass loss drive the evolution of the FG-SG mixing. 
Despite the initial differences in the SG concentration, all systems
approach a state of complete mixing ($\rhratio\sim 1$) after
approximately the same number of $T_{1/2}$ and $T_{tid}$ time scales,
and after losing 60--70 percent of their mass.

Fig.~\ref{fig:ndepend} compares the evolution of $\rhratio$ as a
function of $t/T_{1/2}$ for clusters with the structure of the $\rten$
system but different values of $N$.  The time evolution of $\rhratio$
shows only a very weak dependence on $N$.
\subsection{Connection between SG/FG mixing and $t/t_{rh}(t)$}
While, as pointed out at the beginning of this section, the goal of the simulations presented in this paper is to understand the fundamental aspects of multiple-population cluster dynamics and not to directly compare the results of simulations with observations, we nonetheless think
it is important to identify those parameters that can be most easily
measured for real clusters and might be correlated with the degree of
SG/FG mixing.

One observable quantity that can be reliably estimated for all
clusters is the current half-mass relaxation time, $t_{rh}(t)$.  In
Fig.~\ref{fig:rhrall2} we show the evolution of the degree of mixing
(as measured by $\rhratio$) versus the ratio $t/t_{rh}(t)$.  This
figure suggests that $t/t_{rh}(t)$ may be a good observational
indicator of the extent to which a cluster still retains memory of the
initial SG segregation.

It is important to recognize from Fig.~\ref{fig:rhrall2} that, when interpreting cluster-to-cluster 
differences in the current SG--FG mixing state, one must also consider
possible differences in the initial degree of SG concentration.  For
example, two clusters starting with initial conditions similar to our
{\rtwentyf} and {\rfive} simulations might be characterized by
different relative SG--FG spatial distributions even if observed at
the same dynamical phase and after having lost the same amount of mass
(see Fig.~\ref{fig:rhrall}).  Observations of clusters thought to be
at the same dynamical phase may thus shed light on the extent of the
differences in the initial structural properties and concentration of
the SG subsystem.

\subsection{Spatial mixing and the radial variation of the FG-SG
  number ratio}
\label{sec:radial}
As shown in Figs. \ref{fig:lagrfig}, the FG--SG
mixing process occurs more efficiently and on a shorter time scale in
the cluster inner regions, where the local two-body relaxation time
scale is shorter.  In this section we explore further the evolution of
the radial dependence of the mixing process.

\begin{figure}    
\centering{
\includegraphics[width=7.5cm]{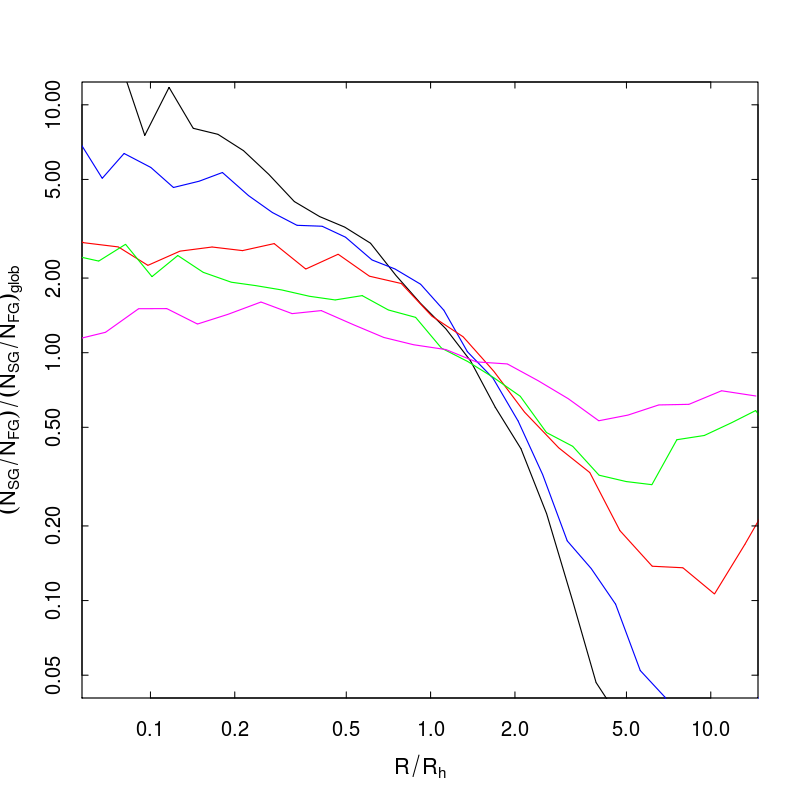}
}
\centering{
\includegraphics[width=7.5cm]{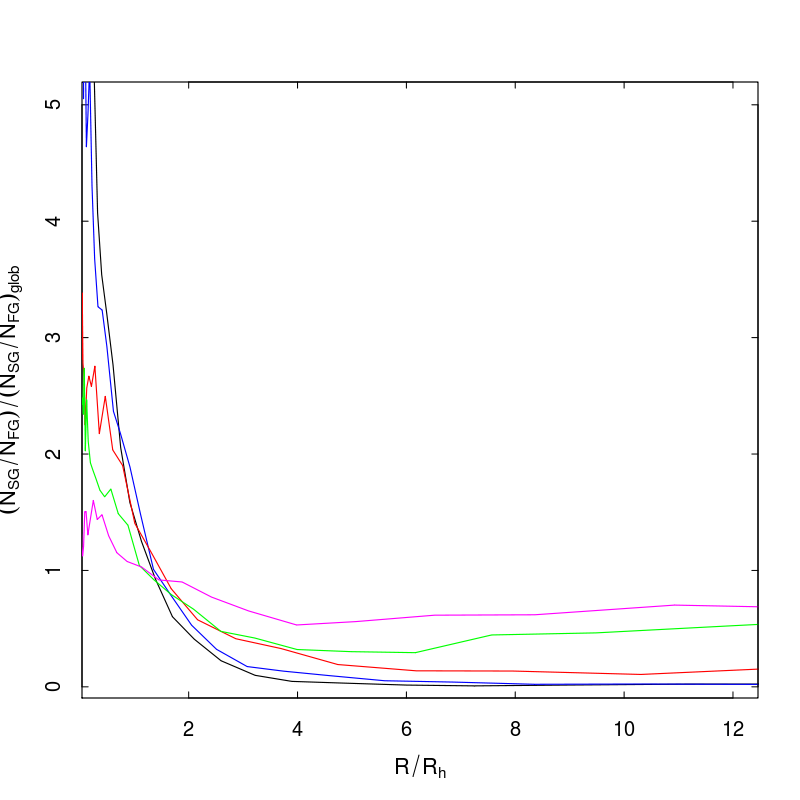}
}
\caption{Evolution of the radial profile of $\nratio$ for the {\rten}
  system.  Radius is normalized to the instantaneous half-mass radius,
  and $\nratio$ is normalized to the global SG-to-FG number ratio.  In
  each panel the profiles are shown at $t/t_{rh}(0)$ [and $t/T_{1/2}$,
  $t/t_{rh}(t)$] approximately equal to $0.2~(0.003,0.2)$ (black),
  $1~(0.015,1)$ (blue), $5~(0.08,3.5)$ (red), $15~(0.23,5)$ (green
  line), and $45~(0.7,10)$ (magenta).  The top panel shows the profiles
  with a log-log scale to highlight the inner structure while the
  lower panel shows the same profiles in linear scale to highlight the
  outer regions.}
\label{fig:nrprofile}
\end{figure}

Fig.~\ref{fig:nrprofile} shows the radial profiles of the number ratio
of SG to FG stars $\nratio$ for the {\rten} system at different stages
of its evolution (the evolution of the radial profile shown in this
figure is representative of the other systems investigated).  The
system starts with the SG population concentrated in the innermost
regions; as the cluster evolves and the FG and SG stars mix, the inner
flat portion of the $\nratio$ profiles progressively extends toward
the outer regions.  Complete mixing corresponds to a flat profile
extending over the entire cluster.

During the evolution the $\nratio$ radial profile is characterized by
an approximately flat inner region followed by a declining outer portion
corresponding to regions increasingly dominated by FG stars.  In the
outer regions, the $\nratio$ profile is characterized again by an
approximately flat portion followed, in some cases, by a weak final rise in
the profile in the cluster outermost regions. 

As our simulations show, during a large fraction of a cluster
evolution, the SG and the FG populations are not completely mixed and
$\nratio$ varies with the distance from the cluster center.  This fact
has several implications: First, some memory of the initial SG
segregation predicted by D'Ercole et al. (2008) should still be
preserved and observable in many clusters today.  Second, the
observational determination of the SG-to-FG number ratio at a given
distance from the center of a cluster will, in general, differ from
the global value of that quantity [hereafter we will refer to the
global value of $\nratio$ as $\nratioglob$].  Third, when exploring
cluster-to-cluster differences in the SG-to-FG number ratio, we must take into account the possibility that part of these differerences might arise if observations cover
different radial zones of different clusters, relative to the
half-mass radius.

The global SG-to-FG number ratio and its cluster-to-cluster variation are key
ingredients of, and constraints on, our ability to understand and
model the formation and evolution of multiple populations in globular
clusters.  It is therefore important to properly understand the relationship between the observational estimates of the SG-to-FG number ratio and its actual global value. 

\begin{figure}    
\centering{
\includegraphics[width=8.5cm]{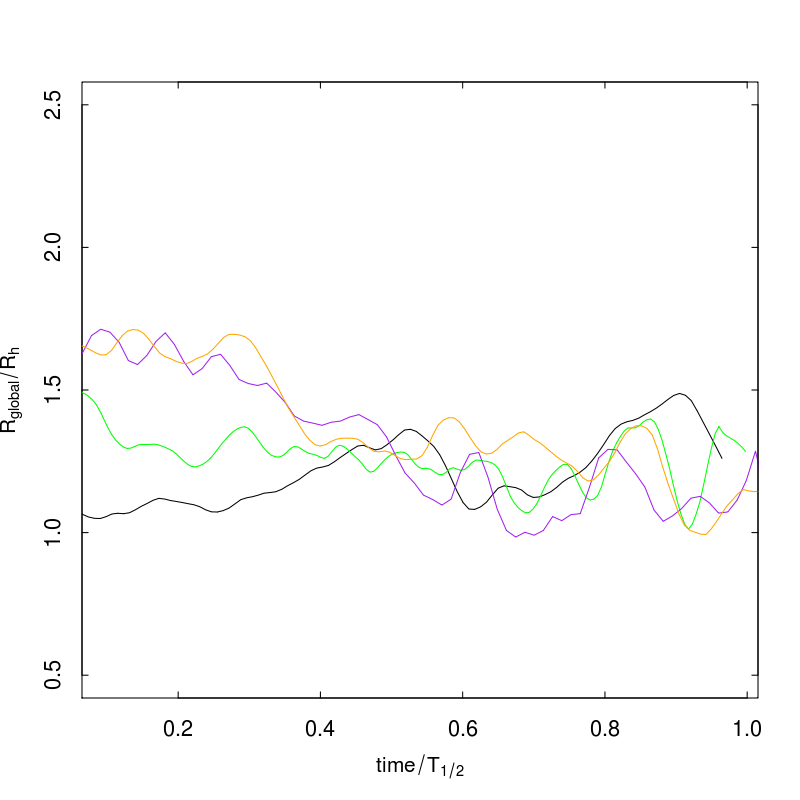}
}
\caption{Time evolution of the ratio of the radius $R_{glob}$, at
  which the local value of $\nratio$ equals the global value
  $\nratioglob$, to the cluster half-mass radius, $R_h$.  Black, green, purple and orange lines refer, 
  respectively, to the \rtwentyf, \rten, \rfive, and {\rtwopf}
  simulations.}
\label{fig:rglob}
\end{figure}

Fig.~\ref{fig:rglob} shows the time evolution of the radius,
$R_{glob}$, at which the local value of $\nratio$ equals the global
value $\nratioglob$. The evolution of $R_{glob}$ is shown only until
$t/T_{1/2} \sim 1$ since, beyond this time, as the system approaches
complete mixing $R_{glob}$ becomes very noisy and not well defined.
Fig.~\ref{fig:rglob} shows that $\nratio$ generally equals
$\nratioglob$ at radii $\sim (1-2) R_{h}$.  These results provide a
first approximate indication of the range of radial distances to be
targeted in observational studies in order to estimate the global
SG-to-FG number ratio from observations covering only a limited radial
range.  Additional simulations with a spectrum of masses are currently
underway, and will allow a refined estimate of $R_{glob}$ and more
direct comparisons with observations.

\subsection{Evolution of the global SG-to-FG number ratio}

As shown by D'Ercole et al. (2008), the transition from a cluster
initially dominated by FG stars to one with similar numbers of SG and
FG stars (or even dominated by SG stars) occurs mainly during the
cluster's early evolution. This transition is due to the loss of FG stars during
the cluster's early expansion triggered by the loss of SNII ejecta. 

\begin{figure}    
\centering{
\includegraphics[width=8.5cm]{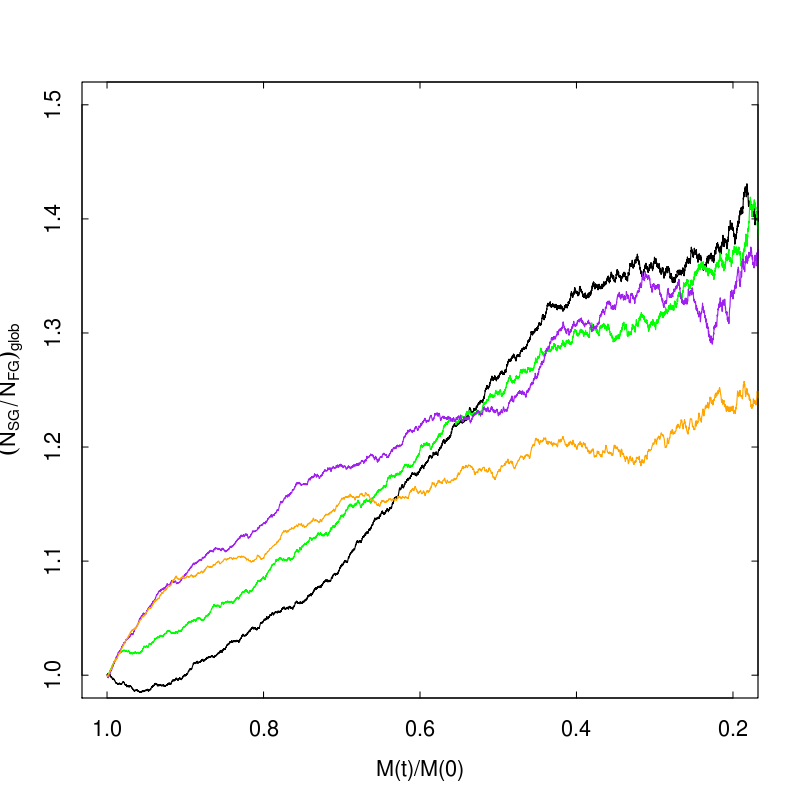}
}
\caption{Time evolution of $\nratioglob$ for the simulations
  {\rtwentyf} (black), {\rten}(green), {\rfive}(purple), {\rtwopf}
  (orange).}
\label{fig:nsgevol}
\end{figure}

During the subsequent long-term evolution, mass loss due to two-body
relaxation removes both FG and SG stars from the cluster.  Until the
two populations are completely mixed, two-body relaxation will still
lead to a slight preferential loss of FG stars and further increase
the SG-to-FG number ratio (although this increase is much smaller than
that during the cluster early evolution; see D'Ercole et al. 2008).
Fig.~\ref{fig:nsgevol} shows the time evolution of the global SG-to-FG
number ratio, $\nratioglob$.  For the systems studied here, during the
long-term evolution $\nratioglob$ increases by only a factor of $\sim
1.3-1.4$.

If stars with chemical properties typical of observed SG populations
can form only in globular clusters, the SG stars observed in the
Galactic halo (see e.g. Martell \& Grebel 2010, Carretta et al. 2010b)
must have been lost by clusters mainly during the long-term evolution
phase explored in this paper.  As discussed by Vesperini et al. (2010;
see also Carretta et al. 2010b, Schaerer \& Charbonnel 2011), the
fraction of SG stars in the Galactic halo is a very important quantity
that can shed light on the connection between globular clusters, their
formation and dynamical history, and their contribution to the
assembly of the Galactic halo.  

\section{Discussion and Conclusions} 
Our previous study of the formation and evolution of multiple
populations (D'Ercole et al. 2008) showed that SG stars form in the
innermost regions of a cluster initially dominated by FG stars.
Following the early cluster expansion triggered by mass loss due to
SNII, a large fraction of FG
stars escape the cluster, leaving a system with comparable numbers of
FG and SG stars, with the SG stars strongly segregated near the
cluster center.  The simulations presented here follow the subsequent
long-term evolution of these clusters, driven by two-body relaxation.
By means of a survey of N-body simulations, we have investigated the
dynamical evolution of multiple-population clusters, focusing our
attention on cluster structural evolution and the FG-SG mixing
process.  We have studied initial conditions characterized by
different degrees of the initial concentration of the SG subsystem.

Two-body relaxation is the key process driving the mixing of the two
populations.  As the local two-body relaxation time scale increases in
the cluster outer regions, this process becomes less efficient at
larger distances from the cluster center.  Moreover, as soon as an
energy source in the cluster core 
causes a cluster to start expanding, the cluster relaxation time
increases with time and the rate of its dynamical aging increasingly
slows down.  A significant slowdown in the evolution toward complete
mixing ensues (see Figs. \ref{fig:rhr25} and \ref{fig:tau}).

Our simulations show that as cluster evolution continues,
relaxation-driven mass loss affects the mixing process by slowing and
eventually reverting the growth of the cluster relaxation time, and by
causing the loss of the cluster's outer unmixed layers.  By comparing
the evolution of clusters with the same initial structure but
different mass-loss rates, we have quantified this process and
illustrated the joint role of internal relaxation and tidal mass loss
in the evolution toward complete mixing.  Our simulations imply that,
unless a cluster has lost a significant fraction of its mass ($\simgt
60-70 \%$; as discussed in the paper, we refer here to the mass lost
during the cluster long-term evolution, and not to the early loss of
FG stars), FG-SG mixing will not be complete and
the SG stars will be centrally concentrated relative to
the FG population (see Fig.~\ref{fig:rhrall}).

To further explore the radial dependence of the mixing process, we
have followed the time evolution of the radial profile of the SG-to-FG
number ratio, $\nratio$ (see Fig.~\ref{fig:nrprofile}).  We have shown
that as SG--FG mixing proceeds, the $\nratio$ profile can be divided
into three different regions: (1) an inner
region where $\nratio$ is flat; (2) an intermediate region where
 the FG is increasingly dominant, with
$\nratio$ declining with increasing radius; (3) an outer region where
the $\nratio$ profile flattens (the outermost cluster regions are in some cases characterized by a slightly rising $\nratio$ profile).

Since, as discussed above, reaching complete mixing (and therefore a
flat $\nratio$ radial profile) requires the cluster to be in an
advanced phase of dynamical evolution, with the loss of a significant
fraction of stars, we expect that a $\nratio$ profile characterized by
the regions described above will be found in all clusters which have
not undergone strong relaxation-driven mass loss---possibly a large
fraction of the Galactic globular cluster population.  Although a
detailed and quantitative comparison with observations is beyond the
scope of the analysis carried out in this paper, our study provides a
preliminary indication of the link between the level of SG/FG mixing
and parameters that can be determined observationally for real
clusters.  Specifically, we have illustrated the relation between
SG/FG mixing and the ratio $t/t_{rh}(t)$ (see
Fig.\ref{fig:rhrall2}).

A few observational studies have explored the differences in the
spatial distribution of FG and SG stars in globular clusters, and find
that, in agreement with the prediction of the formation and evolution
models presented in D'Ercole et al. (2008) and the results presented
in this paper, SG stars do tend to be concentrated in the cluster
inner regions (Bellini et al. 2009, Carretta et al. 2010a, Lardo et
al. 2011, Kravtsov et al. 2010, 2011 Nataf et al. 2011, Johnson \&
Pilachowski 2012, Milone et al. 2012).  It is interesting to note the
similarity between the overall shape of the $\nratio$ profile found in
our simulations and 
the observational profiles as presented in Lardo et al. (2011),
Bellini et al. (2009), and Milone et al. (2012). It is interesting to
point out that all these clusters have values of $t/t_{rh}(t)$
(assuming a common age of $t=11.5$ Gyr and $t_{rh}$ calculated from
the Harris (1996, 2010 edition) catalogue, 
taking into account the approximate relation between the 2D half-mass
radius $R_{h,2D}$, and the 3D half-mass radius $R_h$, $R_{h}\simeq {4\over 3}R_{h,2D}$)
such that the presence of a radial gradient in the $\nratio$
profile is expected, and in agreement with the simulations
presented in this paper (see Fig.\ref{fig:rhrall2}).

Unless a cluster is completely mixed, the SG-to-FG number ratio
determined from observations at a given distance from the cluster
center is in general different from the global value.  This implies
that care must be used in comparing the SG-to-FG number ratio in
different clusters: if different clusters are observed at different
distances from the center, cluster-to-cluster differences in this
number might be found even for systems with the same $\nratioglob$.
Our simulations indicate that, so long as mixing is not complete, the
local value of $\nratio$ measured at $R\approx (1-2) R_h$ is
approximately equal to the global SG-to-FG number ratio.

\section*{Acknowledgments}
E.V. and S.M. were supported in part by grants NASA-NNX10AD86G and
HST-AR-12158.01, HST-AR-12158.05. F.D. and A.D. acknowledge support from grant PRIN-INAF 2011 'Multiple populations in Globular Clusters: their role in the Galaxy assembly'.
We would like to thank M. Bellazzini and C. Lardo for useful discussions.
{}

\end{document}